# Entangled Moiré Chern Insulator in Rhombohedral Graphene

Zaizhe Zhang[1†], Xi Chen[1†], Kenji Watanabe[2], Takashi Taniguchi[3], Zhida Song[1,4,5] and Xiaobo Lu[1,4*]

[1]International Center for Quantum Materials, School of Physics, Peking University, Beijing 100871, China
[2]Research Center for Electronic and Optical Materials, National Institute of Material Sciences, 1-1 Namiki, Tsukuba 305-0044, Japan
[3]Research Center for Materials Nanoarchitectonics, National Institute of Material Sciences, 1-1 Namiki, Tsukuba 305-0044, Japan
[4]Collaborative Innovation Center of Quantum Matter, Beijing 100871, China
[5]Hefei National Laboratory, Hefei 230088, China

† These authors contributed equally to this work.
* Contact author: xiaobolu@pku.edu.cn

**Graphene-based moiré superlattices exhibit novel quantum phenomena driven by pronounced interactions, leading to topological correlated states like orbital Chern insulators exhibiting quantum anomalous Hall effect (QAHE). Typically, intrinsic Chern insulators are stabilized at odd moiré fillings, as even fillings often result in valley-balanced, topologically trivial states at zero magnetic field. In our work, we report the observation of an intrinsic Chern insulator with $C = 1$ state at moiré filling $v = 2$ in rhombohedral octalayer graphene (R8G)/hBN moiré superlattice. Observing such Chern insulators in particular with $C = 1$ at $v = 2$ is intriguing, as each moiré band carries Chern number $C = 1$ or $-1$. We further demonstrate such a state can originate from the entanglement between the low-energy moiré flat bands and high-energy remote bands according to the Hartree-Fock calculation. Our findings extend the known topological phase diagram of rhombohedral multilayer graphene (RMG) moiré systems and establish this platform as highly promising for investigating strong electron correlations and multiband hybridized transport.**

## I. INTRODUCTION

Rhombohedral graphene hosts a variety of topological correlated phases, such as integer QAHE [1–11], fractional QAHE [3–6] and chiral superconductivity [12,13]. However, current investigations of orbital magnetism driven phenomena in RMG/hBN moiré superlattices have been predominantly focused on the low filling factor regime of $|v| \leq 1$ (one electron or hole per moiré unit cell). The topological properties at higher moiré fillings thus remain largely unexplored.

In the other aspect, QAHE or Chern insulator has typically been realized at $|v| = 1$ and $|v| = 3$ in different moiré systems [14–25], corresponding to fully valley-polarized states. Achieving a Chern insulator with valley Chern number $C = 1$ at even integer fillings like $v = 2$ is particularly intriguing and challenging, as such a state could host spin-unpolarized edge modes with each moiré band contributing $C = 1/2$. So far, only anomalous Hall effects have been observed in twisted bilayer graphene [26–28], which is possibly due to proximally induced spin–orbit coupling or partial valley polarization. In this work, we report the observation of a Chern insulator with anomalous Hall resistance near quantized to $R_{xy} = h/e^2$ at $v = 2$ in an R8G/hBN device. We also propose a possible theoretical explanation based on self-consistent Hartree-Fock calculation.

## II. RESULTS

Figure 1(a) illustrate the R8G/hBN device structure. Gate voltages $V_{tg}$ and $V_{bg}$ are applied to the top and bottom graphite gates, respectively, enabling independent tuning of the carrier density $n$ and the electrical displacement field $D$. There is a twist angle of approximately 0.26° between the R8G and the top hBN layer, resulting in a moiré superlattice with a period of about 14.2 nm (Fig. S1).

We measured the longitudinal resistance $R_{xx}$ of the device as a function of $n$ and $D$ at a base temperature of 10 mK (phonon temperature) and zero magnetic field (Fig. S1(a)). The resistance peaks predominantly appear on the electron-doped side ($n > 0$). For $D < 0$ (i.e., when electrons are polarized toward the top layer of R8G, which lies closer to the moiré potential), the electronic states are strongly modulated by the moiré potential, and $R_{xx}$ exhibits a series of equally spaced resistance peaks versus $n$. Based on the relative positions of these peaks, we assign the corresponding moiré filling factors $v = 0, 1, 2, 3$ and $4$, where $v = 4n/n_s$ and $n_s$ is the carrier density corresponding to four electrons per moiré unit cell. In contrast, for $D > 0$, electrons are polarized toward the layer away from the moiré potential, which is the regime of primary interest, and pronounced $R_{xx}$ peaks are observed only at $v = 0, 1$ and $2$.

To probe the regime in which electrons are farther away from the moiré potential ($D > 0$), we performed detailed transport measurements under a small out-of-plane magnetic field $B_z = \pm 0.1$ T. The symmetrized (see Methods) longitudinal resistance $R_{xx}$ and anti-symmetrized Hall resistance $R_{xy}$ are shown in Figs. 1(b) and 1(c), respectively. Consistent with previous reports [3–7,9,10,29,30], pronounced minima in $R_{xx}$ and large values of $R_{xy}$ were observed near $v = 1.09$ and $D/\varepsilon_0 = 0.57$ V/nm, identifying a QAHE state with Chern number $C = +1$ at $v = 1$. We then carried out further magneto-transport measurements of this QAHE state (Figs. 1(d) and 1(e)) The Hall resistance $R_{xy}$ shows a clear sign reversal near zero out-of-plane magnetic field, although it does not fully follow the magnetic field evolution predicted by the Středa formula. Around $B_z = \pm 0.6$ T, the QAHE state was truncated by the arc-shaped resistance ridge states [9], and it reemerges near $B_z = \pm 2.6$ T, indicating a competition between the QAHE state and the resistance ridge states at finite magnetic fields. To further confirm this QAHE state, we performed hysteresis loop measurements (Figs. 1(f) and 1(g)). Both $R_{xx}$ and $R_{xy}$ exhibit significant and robust hysteretic behavior, with $R_{xx}$ dropping below $0.01h/e^2$ at $B_z = 0.15$ T, while $R_{xy}$ is approximately quantized to $h/e^2$. Temperature-dependent hysteresis loop measurements (Fig. 1(h)) indicate a Curie temperature $T_c \approx 1.2$ K for the QAHE state. The inset of Fig. 1(h) shows an Arrhenius plot of $R_{xx}$, from which an energy gap of 0.21 meV is extracted. In addition, Fig. 1(b) reveals prominent $R_{xx}$ peaks corresponding to charge density wave states emerging at fractional fillings $v = 1/3$ and $v = 1/2$, as indicated by the black dashed lines.

Interestingly, at $v = 2$ and for $D/\varepsilon_0$ in the range of 0.85-0.9 V/nm, we observed a pronounced insulating behavior in $R_{xx}$, accompanied by a large anomalous Hall resistance $R_{xy}$ (Figs. 2(a) and (b)). As depicted in Fig. 2(c), a distinct trivial ($C = 0$) correlated insulating state appears at the center of the phase diagram. In the vicinity of this $C = 0$ state, three prominent dips in $R_{xx}$ are observed, and the corresponding $R_{xy}$ values are nearly quantized at $h/e^2$ or $-h/e^2$ (Fig. 2(b)) at $B_z = 0.1$ T. As illustrated schematically in Fig. 2(c), the phases below the $C = 0$ state carry Chern number

$C = -1$, whereas the phase above the $C = 0$ state carries $C = +1$, demonstrating that the chirality of the Chern insulating states at $v = 2$ can be electrically tuned.

To further probe the intrinsic topological properties of this region, we performed a series of magneto-transport measurements (Figs. 2(d-f)). Remarkably, these measurements revealed a $v = 2$ QAHE state with Chern number $C = +1$, which is highly sensitive to the electric displacement field $D$. Figures 2(d) and 2(e) show the symmetrized $R_{xx}$ and anti-symmetrized $R_{xy}$ as a function of $B_z$ and $v$ for several values of $D$ fields. At a larger displacement field ($D/\varepsilon_0 = 0.892$ V/nm), the $C = +1$ QAHE state dominates and persists down to zero magnetic field. As $D$ is reduced, this $C = +1$ Chern insulating state is gradually suppressed, and a trivial insulating phase with $C = 0$ becomes more prominent ($D/\varepsilon_0 = 0.876$ V/nm), accompanied by the gradual emergence of a $C = -1$ Chern insulating state. Upon further decreasing the displacement field, the $C = -1$ Chern insulating phase becomes dominant.

Hysteresis loop measurements at $v = 2.03$ under various displacement fields $D$ (Figs. 2(f)) reveal pronounced hysteretic behavior in $R_{xx}$ across all selected values of $D$. In contrast, $R_{xy}$ exhibits noticeable hysteresis primarily when $C = +1$ state is dominant ($D/\varepsilon_0 = 0.892$ V/nm). Under this condition, $R_{xy}$ approaches the quantized value $h/e^2$ at $B_z = 0.1$ T, while $R_{xx}$ still remains considerably large, likely due to the proximity of the competing $C = 0$ trivial insulating state. At other values of $D$ fields, the dependence of $R_{xy}$ on $B_z$ is nearly linear, with only weak hysteretic behavior.

Given the high sensitivity of the $v = 2$ state to the electrical displacement field, we performed measurements at fixed $v$ while varying $B_z$ and $D$, as shown in Figs. 3(a) and 3(b). Consistent with our earlier analysis, we observed that, as the $D$ field decreases from large to small values, the $v = 2$ state undergoes a sequence of transitions from a $C = +1$ QAHE state to a trivial insulating state with $C = 0$, and finally to a $C = -1$ Chern insulating state. To obtain smaller $R_{xx}$ values for the $C = +1$ QAHE state under finite magnetic fields, we carried out hysteresis loop measurements along the paths indicated by the pink dashed lines in Figs. 3(a) and 3(b). Around $B_z \approx 0.6$ T, $R_{xx}$ approaches a near-zero value ($< 0.01\ h/e^2$), while $R_{xy}$ becomes very close to the quantized value of $h/e^2$ at approximately 0.1 T. Both $R_{xx}$ and $R_{xy}$ displayed clear hysteresis (Figs. 3(d) and (e)). We also extracted line cuts (Figs. 3(f) and 3(g)) for the $C = -1$ Chern insulating state along the cyan dashed lines shown in Figs. 3(a) and 3(b). Similar to the $C = +1$ QAHE state, $R_{xx}$ becomes very small ($< 0.05\ h/e^2$) around $B_z \approx 0.4$ T, and $R_{xy}$ approaches the quantized value of $-h/e^2$ at approximately 0.1 T.

Furthermore, we studied the temperature dependence of the hysteretic behavior of $R_{xy}$ for the $v = 2$ QAHE state. The Curie temperature is determined to be approximately 3.4 K. An Arrhenius plot of $\delta R_{xy} = |\ h/e^2 - R_{xy}\ |$ yields an energy gap of 0.11 meV (Fig. 3(c)). This thermally activated gap is approximately one half of that extracted for the QAHE state characterized by $v = 1$ and Chern number $C = +1$, although the $v = 2$ QAHE state exhibits a slightly higher critical temperature.

To elucidate the physical mechanism underlying the $v = 2$ QAHE state, we propose a theoretical model, as illustrated in Fig. 4. Figure 4(a) shows the phase diagram of the $v = 2$ Hartree-Fock ground states as a function of the interlayer potential energy $V$ and the moiré coupling strength $V_1$. All gapped states are flavor ferromagnetic states, in which both of the two conduction electrons reside in the same valley-spin flavor. This is possible since our system has a large number of layers and a small twist angle, resulting in more than one flat conduction bands in the non-interacting moiré band structure for a wide range of $V$. As shown in Figs. 4(b) and 4(c), under the same interlayer potential $V$ and the same twist angle $\theta$ between hBN and RMG, R8G exhibits significantly flatter bands than R5G, and interactions are expected to induce a much stronger hybridization between the low-energy flat bands and the remote high-energy bands in each valley-spin flavor. Notably, in certain parameter regimes, several lowest conduction bands are almost degenerate, leaving the actually occupied bands as certain hybridizations between them selected by the interaction. The Hartree-Fock calculation then favors flavor ferromagnetic states to minimize the Fock energy. Self-consistent Hartree-Fock calculations show that the two occupied Hartree-Fock bands may have different Chern numbers $C$ = 0 and $C$ = 1 respectively (Fig. 4(d)). As a result, even when the system occupies two conduction bands in one flavor, corresponding to an even filling factor, the total Chern number of the system remains nonzero. This may provide a possible explanation for the absence of a similar half-filling QAHE in previously studied R5G/hBN devices [3,5].

Over a broad range of $V$ and $V_1$, the Hartree-Fock ground state at $\nu = 2$ is found to be gapped with a total Chern number $|C| = 1$, in qualitative agreement with our experimental observation. The chirality of the Chern state can be switched between $C$ = 1 and $C$ = -1 by tuning $V$ even when we keep all ground states to be $K,\uparrow$ polarized (see Methods for details), demonstrating that the chirality can be reversed by tuning the applied electric displacement field. To have a better understanding of such states, we plot the Hartree-Fock band structure of $V = 24 meV, V_1 = 8 meV$ in Fig. 4(d), and find two successive conduction bands with $C = 0, 1$ respectively, the Berry curvature distribution of which are illustrated in Figs. 4(e) and 4(f). The $C = 0$ band has regions with both positive ($\Omega(\mathbf{k}) > 0$) and negative ($\Omega(\mathbf{k}) < 0$) Berry curvature, averaging to zero, while the $C = 1$ band has $\Omega(\mathbf{k}) > 0$ everywhere, contributing to the anomalous Hall effect.

## III. DISCUSSION

Our work demonstrates the experimental realization of a QAHE state at half-filling of a moiré unit cell, substantially extending the known phase diagram of RMG/hBN heterostructures and significantly enriching the established topological framework of moiré superlattice system. These findings highlight RMG as a highly tunable and versatile platform for exploring moiré-induced topological order. Despite these promising observations, a more comprehensive understanding of the underlying mechanisms remains to be developed [31]. Future studies could, for instance, investigate RMG/hBN devices with a larger number of layers to further elucidate the nature of the emergent correlated and topological phases.

## ACKNOWLEDGMENTS

X.L. acknowledges support from the National Key R&D Program (Grant Nos. 2022YFA1403500/02 and 2024YFA1409002) and the National Natural Science Foundation of China (Grant Nos. 12274006 and 12141401). K.W. and T.T. acknowledge support from the JSPS KAKENHI (Grant nos. 21H05233 and 23H02052) and World Premier International Research Center Initiative (WPI), MEXT, Japan.

# APPENDIX: METHODS

## 1. Device fabrication.
The device is fabricated using a standard dry transfer method. Multilayer graphene and hBN flakes are exfoliated on $Si^{++}/SiO_2$ (285 nm) chips. The layer number of graphene flakes was identified through optical contrast. We first identified the rhombohedral region of the multilayer graphene using rapid infrared imaging technique [32], and subsequently confirmed it with Raman spectroscopy, then these rhombohedral domains were precisely isolated into individual regions using femtosecond laser cutting.

A PC (poly bisphenol A carbonate) / PDMS (polydimethylsiloxane) stamp was used to pick up a thin relatively hBN, top graphite gate, top hBN, rhombohedral multilayer graphene, bottom hBN and bottom graphite gate sequentially, the crystalline edges of the rhombohedral graphene and top hBN flakes are meticulously aligned during the transfer process, the whole heterostructure was then released onto a $Si^{++}/SiO_2$ substrate when the PC was backed to 180 °C. Finally, the device was defined into the Hall bar geometry utilizing standard electron beam lithography (EBL) and reactive ion etching (RIE) techniques, and the edge contacts (5 nm Cr / 70 nm Au) were formed using a combination of electron beam evaporation and thermal evaporation techniques.

## 2. Electrical transport measurement.
Unless otherwise specified, all electrical transport measurements were performed within an Oxford Instruments dilution refrigerator, maintaining a base phonon temperature of approximately 10 mK. Keithley 2400 source-meters were used to apply the top gate voltage $V_{tg}$ and bottom gate voltage $V_{bg}$. Stanford Research Systems SR860 lock-in amplifiers were employed to measure the four-terminal longitudinal resistance ($R_{xx}$) and Hall resistance ($R_{xy}$) using an AC current bias ranging from 0.1 to 10 nA at a frequency of 7.777 Hz, and the $R_{xx}$ and $R_{xy}$ were amplified utilizing a Stanford Research Systems SR560 voltage preamplifier.

The carrier density $n$ and electrical displacement field $D$ were defined by $V_{tg}$ and $V_{bg}$, following $n = (C_{tg}V_{tg} + C_{bg}V_{bg}) / e$ and $D = (C_{tg}V_{tg} - C_{bg}V_{bg}) / 2$, where $C_{tg}$ and $C_{bg}$ are the capacitances of the top and bottom gates per unit area, which were calibrated by the quantum Hall effect under high out-of-plane magnetic fields.

## 3. Twist angle extraction.
To determine the twist angle between the top hBN and the rhombohedral multilayer graphene, we use the Brown-Zak quantum oscillations at $D/\varepsilon_0 = 0$ V/nm (Extended Fig. 1). To further verify the twist angle, we utilize the relationship $\lambda = \frac{(1+\delta)a}{\sqrt{2(1+\delta)(1-\cos\theta)+\delta^2}}$ and $n_s = \frac{8}{\sqrt{3}\lambda^2}$, where $\theta$ represents the twist angle of top hBN and top graphene layer of the rhombohedral multilayer graphene, $\delta$ represents the lattice mismatch between graphene and hBN, with a value of approximately 1.7%, $n_s$ denotes the charge carrier density corresponding to the fully filled a superlattice moiré unit cell.

## 4. Symmetrize $R_{xx}$ and anti-symmetrize $R_{xy}$.
Due to imperfect Hall geometry definition in our devices, there is a mixing of the $R_{xx}$ and $R_{xy}$, and we employed the standard procedure of symmetrizing and anti-symmetrizing the raw measured data to correct the mixing effect. This procedure allows us to obtain accurate values of $R_{xx}$ and $R_{xy}$ respectively.

## 5. Self-Consistent Hartree-Fock Calculation at $\nu = 2$.

We first describe the continuum model of R8G without moiré coupling. Around the graphene $K$ point, the Hamiltonian can be expanded as a 16×16 matrix [33]

$$H_K = \begin{pmatrix} v_F \mathbf{k}\cdot\boldsymbol{\sigma} & t^\dagger(\mathbf{k}) & t'^\dagger & \cdots \\ t(\mathbf{k}) & \cdots & \cdots & t'^\dagger \\ t' & \cdots & v_F \mathbf{k}\cdot\boldsymbol{\sigma} & t^\dagger(\mathbf{k}) \\ \cdots & t' & t(\mathbf{k}) & v_F \mathbf{k}\cdot\boldsymbol{\sigma} \end{pmatrix} + H_{ISP} + H_D$$

where $\mathbf{k}$ is measured from the graphene $K$ point $(\frac{4\pi}{3a_G}, 0)$ with $a_G = 0.246$ nm, $v_F$ is the Fermi velocity of single layer graphene, $t(\mathbf{k}) = -\begin{pmatrix} v_4(k_x + ik_y) & -t_1 \\ v_3(k_x - ik_y) & v_4(k_x + ik_y) \end{pmatrix}$ and $t'(\mathbf{k}) = \begin{pmatrix} 0 & 0 \\ t_2 & 0 \end{pmatrix}$ are the coupling between nearest and next nearest graphene layers, $[H_{ISP}]_{ll'} = V_{ISP}\delta_{ll'}|l - \frac{7}{2}|$ is a layer-dependent inversion symmetric chemical potential for layer $l = 0\ldots 7$, and $[H_D]_{ll'} = V\delta_{ll'}(l - \frac{7}{2})$ describes the effect of displacement field $D$. Since the $0^{th}$ layer is adjacent to hBN, $V > 0$ pushes the conduction electrons away from the moiré coupling with hBN. We consider the model parameters $\{v_F, v_3, v_4\} = \{542.1, 34, 34\}$ meV·nm and $\{t_1, t_2, V_{ISP}\} = \{355.16, -7, 16.65\}$ meV [33].

The hBN substrate is aligned adjacent to the top layer ($l = 0$) with twist angle $\theta = 0.26°$. The moiré pattern is characterized by the difference between graphene and hBN $K$ points

$$\mathbf{q}_1 = \frac{4\pi}{3a_G}\left(1 - \frac{R(-\theta)}{1 + \epsilon}\right)\hat{\mathbf{x}}$$

where $\epsilon = 0.01673$ is the lattice mismatch between graphene and hBN, $R(\theta)$ denotes counter-clockwise rotation with angle $\theta$, and $\hat{\mathbf{x}} = (1,0)$. The moiré basis vectors are $\mathbf{g}_1 = \mathbf{q}_2 - \mathbf{q}_3, \mathbf{g}_2 = \mathbf{q}_3 - \mathbf{q}_1, \mathbf{g}_3 = \mathbf{q}_1 - \mathbf{q}_2$, where $\mathbf{q}_j = R\left(\frac{2\pi}{3} * (j-1)\right)\mathbf{q}_1$. The moiré coupling can be well approximated by an effective potential on the $0^{th}$ layer [33–35]

$$V_\xi^K(\mathbf{r}) = V_0 + \left[V_1 e^{i\psi_\xi}\sum_{j=1}^3 e^{i\mathbf{g}_j\cdot\mathbf{r}}\begin{pmatrix} 1 & \omega^{-j} \\ \omega^{j+1} & \omega \end{pmatrix} + h.c.\right]$$

where $\omega = \exp(i\frac{2\pi}{3})$. For simplicity, we neglect the spatially uniform $V_0$ and take $\psi_\xi = 16.65°$. The moiré coupling strength is characterized by $V_1$. Previous studies on RnG with $3 \leq n \leq 7$ estimate $V_1$ to be several meV [33]. Here we use $V_1$ as a tuning parameter. The non-interacting model for $K$' valley can be obtained by time-reversal symmetry.

We consider the dual-gate screened Coulomb interaction $V(\mathbf{q}) = \frac{e^2}{2\epsilon_0\epsilon_r|\mathbf{q}|}\tanh(|\mathbf{q}|d_{sc})$, where $d_{sc} = 10nm$ is the gate distance and $\epsilon_r = 5$ is the relative permittivity.

To avoid double-counting of the interaction energy, we have to choose a "subtraction scheme", which is equivalent to choosing a reference density from which to measure the charge fluctuation, or correctly normal-ordering the interaction Hamiltonian. See Refs. [36,37] for a detailed discussion. Here we choose the "charge neutrality" subtraction scheme with reference density corresponding to occupying all the moiré valence bands, which is widely used in the literature [38–40]. To reduce the computation cost, we project our calculation to the low-energy subspace spanned by the lowest 8 conduction bands and 8 valence bands of the non-interacting model within each valley-spin flavor, i.e. we assume conduction (valence) bands out of this low-energy subspace to be completely empty (occupied). We also impose cutoffs on the Hilbert space $\mathcal{H}$ and interaction $V(\boldsymbol{q})$ at $6|\boldsymbol{q}_1|$ and $4.5|\boldsymbol{q}_1|$ respectively, since states with large momentum from graphene $K$ point cannot be faithfully represented by the continuum model, and only have a weak effect on the low-energy physics.

Our calculation includes both spin and valley flavors. A 12×12 momentum mesh is used to sample the Brillouin zone. Due to the weak spin-orbit coupling of graphene, we allow inter-valley coherence (IVC) states, but assume that $s_z$ is a good quantum number at each lattice momentum. The Hartree-Fock Hamiltonian is uniquely parameterized by the order parameter

$$P_{m\eta,n\eta'}(\boldsymbol{k},s) = \langle c^\dagger_{\boldsymbol{k}m\eta s} c_{\boldsymbol{k}n\eta' s}\rangle$$

where $m, n$ are non-interacting band indices. For each set of parameters $\{V, V_1\}$, we start from 16 different randomly initialized order parameters, including spin-valley polarized and spin-valley balanced states, each with a weak valley-$U(1)$ symmetry breaking term to simulate IVC. The self-consistency loop is terminated when the energy difference between two successive steps is smaller than $0.2~\mu eV$ per unit cell. We then choose the converged state with lowest energy to be the Hartree-Fock ground state. The Berry curvature and Chern number are evaluated from the gauge-invariant Berry flux through each momentum plaquette. When the Berry flux through any plaquette has a larger magnitude than $0.2\pi$, we label "not sure" in the phase diagram, indicating that a finer mesh is needed to determine the Chern number with certainty.

**FIG. 1**

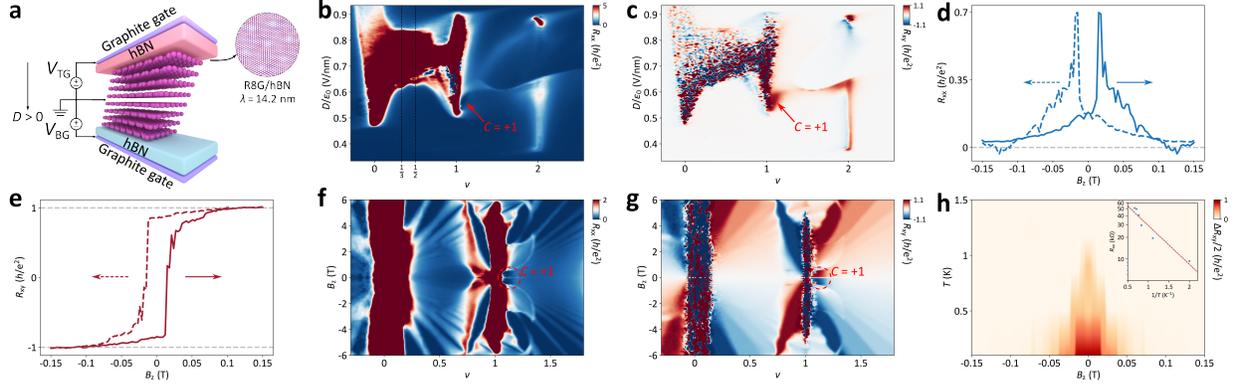

FIG. 1. Phase diagram of R8G/hBN and Chern insulator at $v = 1$. (a) Schematic diagram of the device structure. Due to a relative twist angle of 0.26 degrees and lattice mismatch between R8G and top hBN, a moiré superlattice with a periodicity of 14.2 nm is formed. (b-c) Symmetrized longitudinal resistance $R_{xx}$ and anti-symmetrized Hall resistance $R_{xy}$ versus filling factor $v$ and electrical displacement field $D$ at out-of-plane magnetic field $B_z = 0.1$ T. (d-e) Landau fan diagrams of symmetrized $R_{xx}$ and anti-symmetrized $R_{xy}$ at $D/\varepsilon_0 = 0.57$ V/nm. (f-g) Symmetrized $R_{xx}$ and anti-symmetrized $R_{xy}$ as a function of $B_z$ measured at $v = 1.09$ and $D/\varepsilon_0 = 0.57$ V/nm, dashed and solid lines correspond to sweeping the $B_z$ field back and forth indicated by the arrows. (h) Anomalous Hall effect amplitude $\Delta R_{xy}/2 = |R_{xy} (B_z \text{ sweeping up}) - R_{xy} (B_z \text{ sweeping down})|/2$ versus $B_z$ field and temperature $T$ measured at $v = 1.09$ and $D/\varepsilon_0 = 0.57$ V/nm. The inset in the upper right is extracted from Fig. S2, showing the $R_{xx}$ Arrhenius plot at $B_z = 0.15$ T, with the fitted thermal activation gap value of 0.21 meV. The data in (c-h) are acquired at $T = 10$ mK.

**FIG. 2**

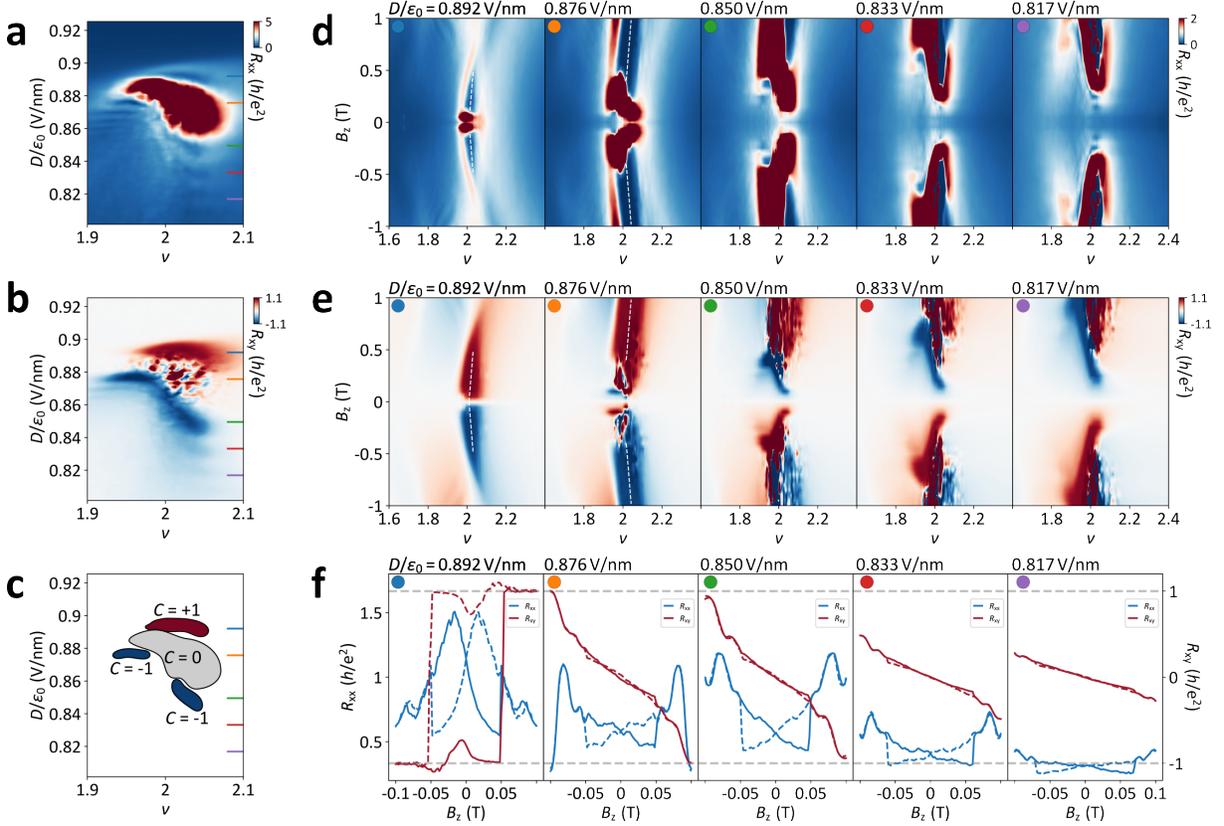

FIG. 2. Competition between multiple strongly correlated phases at selected electrical displacement fields near $v = 2$. (a-b) Zoom-in of the symmetrized $R_{xx}$ and anti-symmetrized Hall resistance $R_{xy}$ as a function of filling factor $v$ and $D$ at $B_z = 0.1$ T. (c) Schematic diagram illustrating the different correlated phases near $v = 2$, extracted from (a-b). Which includes the trivial correlated insulator ($C = 0$) state in the center of the phase diagram, along with the $C = +1$ Chern insulator state above it and the $C = -1$ Chern insulator states below it. (d-e) Symmetrized $R_{xx}$ and anti-symmetrized $R_{xy}$ as a function of $v$ and out-of-plane magnetic field $B_z$ measured at $D/\varepsilon_0 = $ 0.892 V/nm, 0.876 V/nm, 0.850 V/nm, 0.833 V/nm and 0.817 V/nm, respectively (colored solid lines in a-c). The white dashed lines in the figure indicate the evolution of the $C = +1$ Chern insulator state emanating from $v = 2$ with magnetic field according to the Streda formula. (f) Symmetrized $R_{xx}$ and anti-symmetrized $R_{xy}$ as a function of $B_z$, measured at $v = 2.03$ for $D/\varepsilon_0 = $ 0.892 V/nm, 0.850 V/nm, 0.833 V/nm and 0.817 V/nm, and at $v = 1.96$ for $D/\varepsilon_0 = 0.876$ V/nm, respectively. Dashed and solid lines correspond to sweeping the $B_z$ field back and forth indicated by the arrows. All experimental data are acquired at $T = 10$ mK.

**FIG. 3**

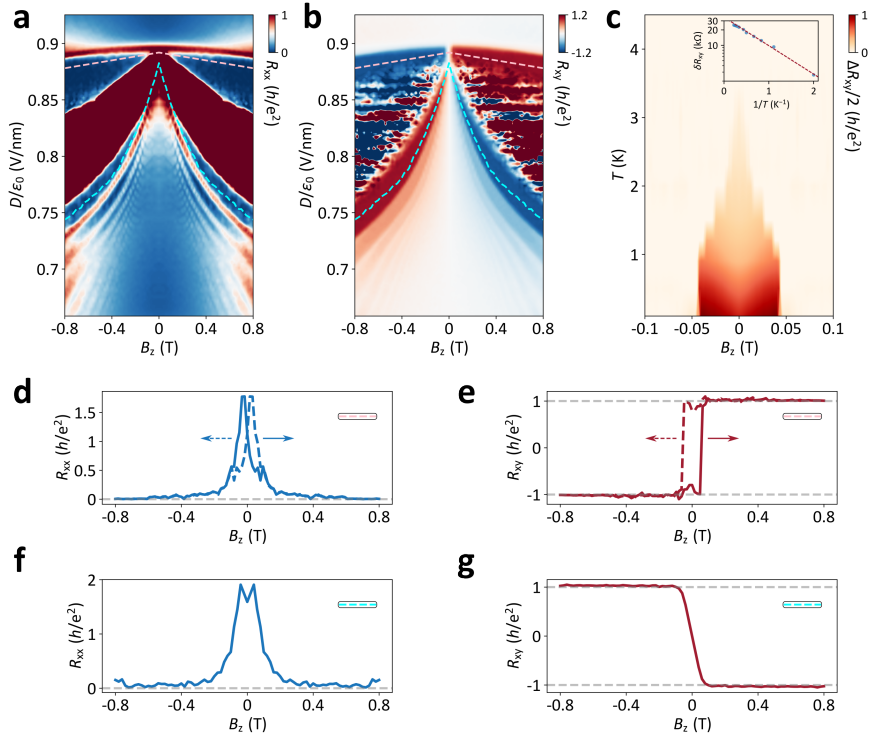

FIG. 3. Tunable-chirality Chern insulators by electrical displacement field $D$ at $v = 2$. (a-b) Symmetrized $R_{xx}$ and anti-symmetrized $R_{xy}$ versus out-of-plane magnetic field $B_z$ and electrical displacement field $D$ at $v = 2.03$ and $T = 10$ mK. (c) Anomalous Hall effect amplitude $\Delta R_{xy}/2 = | R_{xy} (B_z$ sweeping up$) - R_{xy} (B_z$ sweeping down$) | / 2$ versus $B_z$ field and temperature $T$ measured at $v = 2.03$ and $D/\varepsilon_0 = 0.892$ V/nm. The inset in the upper right is extracted from Fig. S2, showing the $\delta R_{xy} = | h/e^2 - R_{xy} (B_z = 0.15$ T$) |$ Arrhenius plot, with the fitted thermal activation gap value of 0.11 meV. (d-e) Symmetrized $R_{xx}$ and anti-symmetrized $R_{xy}$ as a function of $B_z$ measured at $v = 2.03$ and during the evolution of the $B_z$, $D$ is adjusted synchronously along the pink dashed lines in (a-b), dashed and solid lines correspond to sweeping the $B_z$ field back and forth indicated by the arrows, corresponding to a QAHE state characterized by $v = 2$ and $C = +1$. (f-g) Linecuts extracted along the cyan dashed lines in (a-b), corresponding to a Chern insulator state characterized by $v = 2$ and $C = -1$.

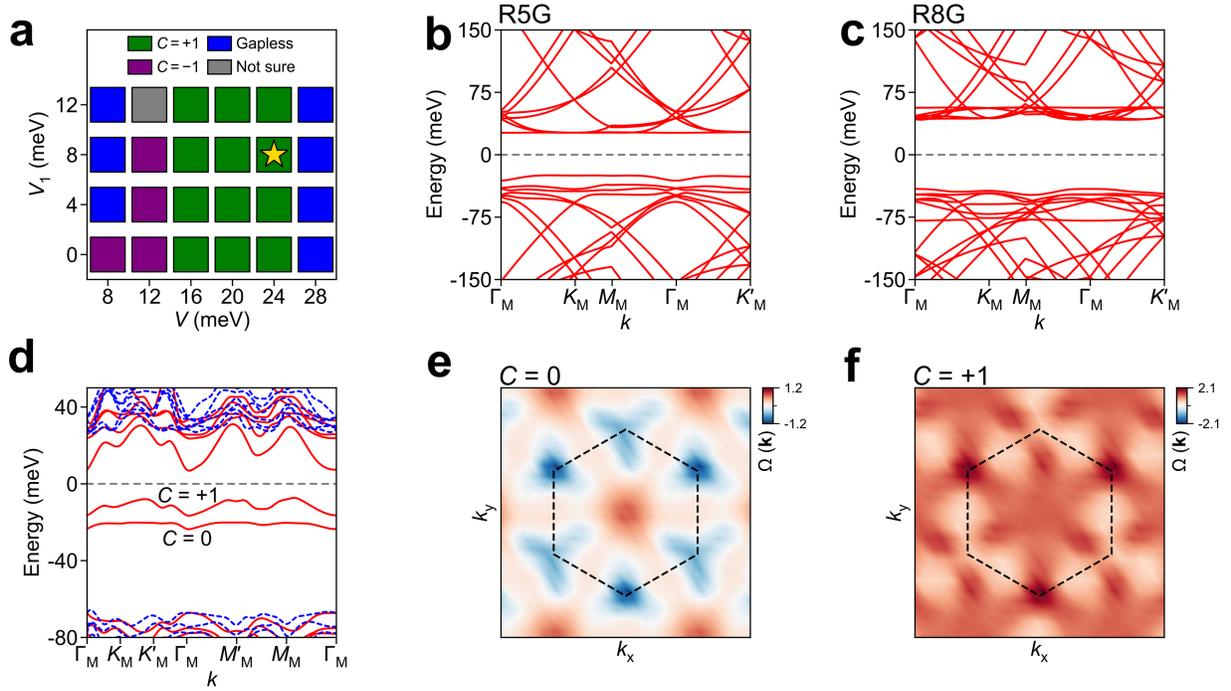

FIG. 4. Theoretical calculation of the QAH state characterized by $v = 2$ and Chern number $C = +1$. (a) Phase diagram of R8G/hBN with $\theta = 0.26°$ at filling factor $v = 2$ as a function of the interlayer potential energy $V$ and the moiré coupling strength $V_1$. The green, purple, blue, and gray regions indicate the $C = +1$ phase, the $C = -1$ phase, the gapless state and the uncertain phase, respectively. (b-c) Non-interacting band structures of the R5G and R8G/hBN single-particle model in $K$ valley with $V = 20$ meV, $V_1 = 8$ meV and $\theta = 0.26°$. (d) Hartree–Fock band structure at the selected parameters $V = 20$ meV, $V_1 = 8$ meV, where two successive conduction bands are obtained and their calculated Chern numbers ($C = 0, 1$) are labeled in the figure. (e-f) Distribution of the Berry curvature for the occupied bands with the Chern number $C = 0$ and $C = +1$. The black dashed lines show the moiré Brillouin zone. The Berry curvature is normalized such that a constant $\Omega (\mathbf{k}) = 1$ corresponds to $C = +1$.